\documentclass[12pt]{article}
\usepackage{epsfig}

\usepackage{amssymb}
\usepackage{amsmath}
\usepackage{amsfonts}

\def\be{\begin{equation}}
\def\ee{\end{equation}}
\def\ba{\begin{array}{c}}
\def\ea{\end{array}}

\def\ben{$$}
\def\een{$$}

\newcommand{\bea}{\begin{eqnarray}}
\newcommand{\eea}{\end{eqnarray}}

\newcommand{\bbr}{\br\!\br}
\newcommand{\kkt}{\kt\!\kt}
\newcommand{\pbr}{\prec\!}
\newcommand{\pkt}{\!\succ\,\,}
\newcommand{\kt}{\rangle}
\newcommand{\br}{\langle}

\newcommand{\brz}{\{\!\{}
\newcommand{\ktz}{\}\!\}}

\newcommand{\ed}{\end{document}}

\begin{document}

\title{Sturm-Schr\"{o}dinger equations: formula for metric}

\author{Miloslav Znojil\\
Nuclear Physics Institute ASCR, \\
250 68 \v{R}e\v{z}, \\
Czech
Republic\footnote{e-mail: znojil@ujf.cas.cz}\\
.
\\
Hendrik B.~Geyer\\
Institute of Theoretical Physics,\\
 University of Stellenbosch
and Stellenbosch Institute for Advanced Study,\\ 7600
Stellenbosch, \\
South Africa\footnote{e-mail: hbg@sun.ac.za}}

\maketitle

\newpage

\abstract{Sturm-Schr\"{o}dinger equations $H\psi=E\,W\psi$ with
$H\neq H^\dagger$ and $W\neq W^\dagger \neq I$ are considered,
with a weak point of the theory lying in the purely numerical
matrix-inversion form of the double-series definition of the
necessary metric operator $\Theta$ in the physical Hilbert space
of states [M. Znojil, J. Phys. A: Math. Theor. 41 (2008) 215304].
This shortcoming is removed here via an amended, single-series
definition of $\Theta$. }

\vspace{2cm}

 keywords:

 {cryptohermiticity,
 Sturm-Schr\"{o}dinger bound-state problem}

\vspace{1cm}


\newpage

\section{The origin of Sturm-Schr\"{o}dinger bound-state equations}

It has already been known to Liouville \cite{Liouville} that after
one pays due attention to boundary conditions an elementary change
of variables may transform a given Sturm-Liouville eigenvalue
problem
 \be
 H^{(1)}\,\psi^{(1)}_n(r)
 = E_n\,
 \psi^{(1)}_n(r)\,,
 \ \ \ \
 H^{(1)}=-\frac{d^2}{dr^2}
 +V^{(1)}(r)
  \label{SErovni}
 \ee
into another Sturm-Liouville eigenvalue problem of just slightly
more complicated form, viz.,
 \be
 H^{(2)}\,\psi^{(2)}_n(x)
 = E_n\,W(x)\,
 \psi^{(2)}_n(x)
 \,,
 \ \ \ \
 H^{(2)}=-\frac{d^2}{dx^2}
 +V^{(2)}(x)\,.
 \label{SErovnice}
 \ee
It is important to notice that both these equations are {\em
isospectral}, i.e.,  we have $E^{(1)}_n = E^{(2)}_n=E_n$ at all $n
= 0, 1, \ldots$. Recently, we employed such an idea in
ref.~\cite{I} (to be called paper I in what follows). We
interpreted there eq.~(\ref{SErovni}) as a standard
Schr\"{o}dinger equation of Quantum Mechanics, with
$H^{(1)}=T+V^{(1)}$ where the differential operator $T$
represented kinetic energy. In parallel, the change of variables
(viz., $r \to x$ and $\psi^{(1)} \to \psi^{(2)}$) has been
prescribed in advance so that we arrived at explicit formulae for
the two new functions $W(x)$ and $V^{(2)}(x)$ entering the new
differential-equation form~(\ref{SErovnice}) of our quantum
bound-state problem. In such a context we found it very natural to
call eq.~(\ref{SErovnice}) a Sturm-Schr\"{o}dinger equation with
Hamiltonian $H$ and weight operator $W$.

The principal advantage of our slightly counterintuitive
transition from eq.~(\ref{SErovni}) to eq.~(\ref{SErovnice})
[notice that we would have $W^{(1)}(r) \equiv 1$ in
eq.~(\ref{SErovni})] has been found in a perceivable
simplification of the physical asymptotic boundary conditions.
This implied that also the practical (say, numerical) solution of
eq.~(\ref{SErovnice}) proved much easier and quicker than the
solution of its predecessor (\ref{SErovni}) (cf. refs.~\cite{I} or
\cite{Wessels} for more details).

\section{The emergence of a nontrivial metric in Sturm-Schr\"{o}dinger
bound-state problems in Quantum Mechanics}

One must be aware that the presence of the nontrivial weight
operator $W(x)$ may make the solution of eq.~(\ref{SErovnice})
less routine. In paper I we paid particular attention to models
(\ref{SErovni}) and (\ref{SErovnice}) where in spite of the
reality of all the spectrum of energies a naive choice of the two
alternative representations ${\cal H}_{(1,2)}^{(F)}$ of the
Hilbert space of states of our quantum system in question happened
to lead to inconsistencies like negative probabilities etc. For
this reason the superscript $^{(F)}$ stands here for ``false".

In paper I we emphasized that the superscript $^{(F)}$ could
equally well stand here for ``friendly". Indeed, once we
temporarily forget about inner products and consider just the
underlying two {\em vector} spaces ${\cal V}_{(1,2)}^{(F)}$ we may
treat them simply as spanned by the respective eigenfunctions or
generalized eigenfunctions $\psi^{(1,2)}_n(\cdot) \equiv \langle
\cdot |\psi^{(1,2)}_n\rangle$ of our respective Hamiltonians.

In such a setting the main difficulty as addressed in paper I was
represented by our specific assumption of the  manifest
non-Hermiticity of our Hamiltonian operators,
 \be
 H^{(1,2)} \neq
 \left [ H^{(1,2)}
 \right ]^\dagger\, \ \ \ \ \  {\rm in }\ \ \ \ \ {\cal
 H}_{}^{(F)}={\cal
 H}_{(1,2)}^{(F)}\,.
  \label{haSErovni}
 \ee
Fortunately, being well aware of the inadequacy of the underlying
choice of the specific and most elementary (often called Dirac's
or Lebesgue's) inner products
 \be
 \langle \psi\,|\,\phi\rangle = \int_{}^{}\,d\mu(r)\, \psi^*(r) \phi(r)
 \,,
 \ \ \ \ \ \ |\psi\kt\,,|\phi\kt \in {\cal H}_{}^{(F)}
 \,
 \ee
we immediately recalled the well known recipe as described, e.g.,
in  refs.~\cite{Geyer} or \cite{SIGMA}.

Let us briefly recollect the mathematical core of the recipe which
lies in a redefinition of the operation of Hermitian conjugation.
Technically this means that in the first step we decide to treat
our Hilbert space ${\cal H}_{}^{(F)}$, formally, as a vector space
${\cal V}_{}^{(F)}$ of kets $|\psi\kt$ complemented by the
``usual" definition of the dual space $\left [{\cal
V}_{}^{(F)}\right ]'$ (marked by a prime) of bras $\br \psi|$. In
this setting the operation of Hermitian conjugation ${\cal
T}^{(general)}$ is, in general, non-unique \cite{Messiah}. Still,
people usually ignore this ambiguity and decide to work in full
analogy with the linear algebra and finite-dimensional Hilbert
spaces. Thus, they specify the dual vector space of linear
functionals, i.e., bra vectors as quantities constructed from
their ket-vector partners by the very specific Dirac's operation
${\cal T}^{(Dirac)}$ of transposition plus complex conjugation.

In the second step we replace the above-mentioned Dirac's
definition of conjugation in Hilbert space ${\cal H}_{}^{(F)}$,
viz.,
 \be
 \br \psi|:={\cal T}^{(Dirac)}\,|\psi\kt\,
 \ \ \ \ \ \ {\rm in } \ \ \ \ \ {\cal H}_{}^{(F)}
 \label{jedub}
 \ee
by  a non-Dirac definition of conjugation which amounts to an
introduction of another Hilbert space ${\cal H}_{}^{(S)}$. This
new, unitarily non-equivalent (i.e., inner product non-preserving)
representation of Hilbert space is to be formed by the {\em same}
vector space ${\cal V}_{}^{(S)} \ \equiv\ {\cal V}_{}^{(F)}$ and
by a {\em different} ``primed" vector space of linear functionals
$\left [{\cal V}_{}^{(S)}\right ]' \ \neq\ \left [{\cal
V}_{}^{(F)}\right ]'$.

On the level of notation we {\em must} graphically distinguish
between the dual elements  $\br \psi| \in \left [{\cal
V}_{}^{(F)}\right ]'$ given by eq.~(\ref{jedub}) and, in our
notation \cite{I}, elements $\bbr \psi| \in \left [{\cal
V}_{}^{(S)}\right ]'$ defined by the {\em different} relation
 \ben
 \bbr \psi|
 \ \equiv\ \br \psi|\,\Theta
 :={\cal T}^{(non-Dirac)}\,|\psi\kt\,
 \ \ \ \ \ \ {\rm in } \ \ \ \ \ {\cal H}_{}^{(S)}\,,
 \ \ \ \ \ \ \Theta \neq I
 \een
or, in the coordinate representation,
 \be
 \bbr \psi\,|\cdot\kt =
 \int_{}\,d\mu(r)\,\psi^*(r)\,\Theta(r,\cdot)\
 \,.
 \ee
This redefinition realizes the transition from the false Hilbert
spaces  ${\cal H}_{(1,2)}^{(F)}$ to their unitarily non-equivalent
alternatives ${\cal H}_{(1,2)}^{(S)}$ where the superscript
$^{(S)}$ stands for ``standard". In the coordinate represenation
one speaks about the modified, double-integral definition of the
inner product
 \be
 \bbr \psi\,|\phi\kt =
 \int_{}\,d\mu(x)\,
 \int_{}\,d\mu(y)\, \ \psi^*(x)\,
 \Theta(x,y)\,\phi(y) \,
 \label{convention}
 \ee
applicable to any two wave functions $\psi$ and $\phi$ in ${\cal
H}_{(1,2)}^{(S)}$.


\section{Generalized Dyson-type mappings $\Omega$
\label{fff.}}

The key purpose of introducing the  above-described nontrivial
metrics $\Theta=\Theta_{(1,2)}$ lies in making our Hamiltonians
self-adjoint in ${\cal H}_{(1,2)}^{(S)}$.
%
%
%
For our further purposes we shall drop the subscripts and restrict
our attention just to the Sturmian cases with subscripts $_2$.
Next we introduce an invertible map $\Omega: {\cal V} \to {\cal
A}$ connecting our vector space of functions (with ket elements
$|\phi\kt \in {\cal V}$) with another, abstract vector space
${\cal A}$ composed of certain not yet specified spiked-ket
elements,
 \be
 |\psi \pkt = \Omega\,|\psi \kt\,,
 \ \ \ \
 |\psi \kt \in {\cal V}\,,
 \ \ \ \
 |\psi \pkt \in {\cal A}\,.
 \label{elemer}
 \ee
Using such a notation we shall assume the solvability of the {\em
non-selfadjoint doublet} of the Sturm-Schr\"{o}dinger equations
composed of eq.~(\ref{SErovnice}) and of its dual defined in the
sense of space ${\cal H}^{(F)}$, i.e., in the same vector space
${\cal V}$. Thus, in an obvious shorthand notation we shall
consider the following two equations,
 \be
 H^{}\,|\,{\lambda}\rangle
 =\lambda\,W\,|\,{\lambda}\rangle\,
 \label{sturmjed}
 \,,\ \ \ \ \ \ \ \ \
 H^\dagger \,|\,{\lambda'}\rangle\!\rangle
 =\lambda'\,W^\dagger\,|\,{\lambda'}\rangle\!\rangle\,
  \ee
By assumption, they define two families of elements of vector
space ${\cal V}$.
%
%
After the application of mapping $\Omega$ one also obtains the
corresponding families of elements of the abstract vector space
${\cal A}$. Working here, exclusively and solely, with the
standard and usual Dirac's conjugation we may introduce also the
dual space ${\cal A}'$ and a related third, abstract Hilbert space
${\cal H}^{(P)}$ where the superscript $^{(P)}$ stands for
``physical".

These conventions lead to several immediate consequences. Firstly,
without any danger of confusion, the elements of the dual space of
linear functionals may simply be denoted by the spiked ket
symbols,
 \be
  \pbr\,\psi\,| = \langle\,\psi\,|\,
 \Omega^\dagger\,\in\,
 {\cal A}'\,.
 \label{lumpacif}
 \ee
Secondly, the lower-case isospectral equivalent $h =
\Omega\,H\,\Omega^{-1}$ of our original non-Hermitian upper-case
Hamiltonians $H\neq H^\dagger$ as well as the parallel partner $w
= \Omega\,W\,\Omega^{-1}$ of any original non-Hermitian specific
``weight" operator $W\neq W^\dagger$ may and will be assumed
self-adjoint in the corresponding abstract physical Hilbert space
${\cal H}^{(P)}$. In a way explained in ref.~\cite{SIGMA} the
latter space is  specified by its spiked-ket elements
(\ref{elemer}) and by its spiked-bra linear functionals
(\ref{lumpacif}).

All this means that we must require that
 \ben
 h^\dagger = \left (\Omega^{-1}
 \right )^\dagger\,H^\dagger\,\Omega^\dagger= h\,,
 \ \ \ \ \ \ \ \ \ \
 w^\dagger = \left (\Omega^{-1}
 \right )^\dagger\,W^\dagger\,\Omega^\dagger= w\,,
 \een
or, after a trivial re-arrangement,
 \be
 H^\dagger = \Theta\,H\,\Theta^{-1}\,,
 \ \ \ \ \
 W^\dagger = \Theta\,W\,\Theta^{-1}\,
 \label{prope}
 \ee
where we abbreviated $\Theta = \Omega^\dagger\,\Omega$. This shows
how the concept of Hermiticity in physical Hilbert space ${\cal
H}^{(P)}$ remains equivalent to the apparent non- or
pseudo-Hermiticity  or cryptohermiticity in ${\cal H}^{(F)}$. In
the language of ref.~\cite{Geyer}, property (\ref{prope}) of all
our operators should be called quasi-Hermiticity in space ${\cal
H}^{(S)}$ where the metric is nontrivial.


The pull-back of all the theory to the abstract Hilbert space
${\cal H}^{(P)}$ clarifies that all our present considerations
need not leave the domain covered by standard textbooks
\cite{Messiah}. In particular, due to the assumption of
invertibility of the mappings $\Omega$ we may replace our two
original upper-case Sturmian problems (\ref{sturmjed}) by their
common lower-case reincarnation
 \be
 h^{}\,|\,{\lambda}\pkt
 =\lambda\,w\,|\,{\lambda}\pkt\,
 \label{lcsturmjed}
 \ee
which is necessarily self-adjoint  in ${\cal H}^{(P)}$. Its
simplicity facilitates the derivation of the Sturmian
orthogonality relations
 \be
 \pbr \lambda\,|\,w\,|\,\lambda'\pkt\ =\
 \pbr \lambda\,|\,w\,|\,\lambda\pkt\,\cdot\,
 \delta_{\lambda,\lambda'}\,
 \label{luma}
 \ee
and of the Sturmian completeness relations,
 \be
 I = \sum_{\lambda}\,|\,\lambda\pkt
 \,
 \frac{1}{\pbr \lambda\,|\,w\,|\,\lambda\pkt}
 \,\pbr \lambda\,|\,w\,
 \ee
as well as of the Sturmian spectral representation
 \be
 h = \sum_{\lambda}\,w\,|\,\lambda\pkt\,
 \frac{\lambda}{\pbr \lambda\,|\,w\,|\,\lambda\pkt}
 \,\pbr \lambda\,|\,w\,
 \ee
of the lower-case Hamiltonian in ${\cal H}^{(P)}$.

\section{Biorthogonal bases and the double-series formula for the metric
\label{ddd.}}

Our original Hilbert space ${\cal H}^{(F)}$ is, by assumption, so
simple that it makes sense to transfer all the above formulae to
this space. Thus, the insertion of definitions (\ref{elemer}) and
(\ref{lumpacif}) in eq.~(\ref{luma}) yields the orthogonality
relations in ${\cal H}^{(F)}$,
 \be
 \langle\,\lambda\,|\,
 \Omega^\dagger\,w\,\Omega\,|\,\lambda'\,\kt =
 \langle\,\lambda\,|\,\Theta\,W
 \,|\,\lambda'\,\kt =
 \langle\,\lambda\,|\,\Theta\,W
 \,|\,\lambda\,\kt\,\cdot\,
 \delta_{\lambda,\lambda'}\,.
 \label{deluma}
 \ee
Similarly, the appropriately adapted version of the completeness
is obtained,
 \be
 I = \sum_{\lambda}\,|\,\lambda\,\kt \,
 \frac{1}{
 \langle\,\lambda\,|\,\Theta\,W
 \,|\,\lambda\,\kt}
 \,
 \langle\,\lambda\,|\,
 \Theta\,W\,.
 \ee
Finally, the spectral decomposition of the Hamiltonian acquires
the following form,
 \be
 H = \sum_{\lambda}\,W\,|\,\lambda\,\kt
 \,
 \frac{\lambda}{
 \langle\,\lambda\,|\,\Theta\,W
 \,|\,\lambda\,\kt}
 \,\langle\,\lambda\,|\,
 \Theta\, W\,.
 \ee
Moreover, the double-ket eigenstates $
|\,\lambda\,\rangle\!\rangle$ of $H^\dagger$ may be understood as
equal to the products $\Theta|\,\lambda\,\rangle$ \cite{I,SIGMA}.

The key benefit of our return to the metric-independent and
computation-friendly space ${\cal H}^{(F)}$ is that we may
evaluate and set the matrix elements
$\langle\,\lambda\,|\,\Theta\,W \,|\,\lambda\,\kt$ in
eq.~(\ref{deluma}) equal to one via a suitable normalization of
the basis \cite{SIGMA}. This convention will perceivably simplify
our formulae. The simplification would remain applicable whenever
the Hermitian product $\Theta\,W$ stays positive definite (which
is to be assumed). The first consequence of the resulting
simplification of the formulae is that we may introduce the
further, curly-ket-marked wave-function vectors and their
curly-ketket partner vectors defined, respectively, as follows,
 \ben
 |\,\psi\,\} = W\,|\,\psi\,\kt\,,
 \ \ \ \ \
 |\,\psi\,\ktz = W^\dagger\,|\,\psi\,\kkt\,.
 \een
In terms of these new abbreviations we may stay working inside the
most comfortable Hilbert space ${\cal H}^{(F)}$ and write down the
following alternative and optional forms of the orthogonality
conditions,
 \be
 \langle\,\lambda\,|\,\Theta\,W
 \,|\,\lambda'\,\kt=
 \brz\,\lambda\,|
 \,\lambda'\,\kt=
 \bbr\,\lambda\,|\,\lambda'\,\}=
 \delta_{\lambda,\lambda'}
 \,.
 \ee
Similarly we may derive the two alternative forms of the
completeness relations,
 \be
 I = \sum_{\lambda}\,|\,\lambda\,\kt \,
  \brz\,\lambda\,| = \sum_{\lambda}\,|\,\lambda\,\} \,
  \bbr\,\lambda\,|
 \,.
 \label{copu}
 \ee
Finally, the menu of several alternative spectral-representation
expansions can be replaced by the more or less unique, most
compact expressions
 \be
 W = \sum_{\lambda}\,|\,\lambda\,\} \,
  \brz\,\lambda\,|
 \,,\ \ \ \ \ \ \ \ \ \
 H = \sum_{\lambda}\,|\,\lambda\,\}
 \,
 {\lambda}
 \,\brz\,\lambda\,|\,
 \label{forumro}
 \ee
representing the weights and the Hamiltonian operators.

At this moment one could decide to employ the method of
ref.~\cite{I} and to insert spectral formulae (\ref{forumro}) in
the fundamental quasi-Hermiticity constraint or equation
$H^\dagger\,\Theta= \Theta\,H$ yielding
 \be
 \sum_{\lambda}\,|\,\lambda\,\ktz \,\lambda\,\{\lambda|\,\Theta=
  \sum_{\lambda}\,\Theta\,|\,\lambda\,\} \,\lambda\,\brz
  \lambda|\,.
  \ee
This relation would strongly suggest that the most natural
Sturmian analogue of the well known single-series $W=I$ formula
should be sought via the double series ansatz
 \be
 \Theta = \sum_{\lambda,\lambda'}\,|\,\lambda\,\ktz \,
 M_{\lambda,\lambda'}\,
 \brz\,\lambda'\,|
 \,,\ \ \ \ \ \ \ \ \ \
 M_{\lambda,\lambda'} =
 \br\,\lambda\,|\,\Theta\,|\,\lambda'\kt
 \,.
 \label{reforum}
 \ee
This approach has been accepted in paper I and contributed to its
discouraging conclusions. Fortunately, a better strategy exists.
Its core and main consequences will now be described in our last
section \ref{5.s}.

\section{The single-series formula for $\Theta=\Theta(H,W)$
\label{5.s}}

We arrived at the heart of our present message. The identity
$|\psi\kkt = \Theta\,|\psi\kt$ inspires us to start from
eq.~(\ref{copu}) and to obtain the following unexpected but
sufficiently simple result
 \be
 \Theta=  \sum_{\lambda}\,|\,\lambda\,\kkt \,
  \brz\,\lambda\,|\,.
   \label{neebe}
 \ee
The impression of an apparent non-Hermiticity of this asymmetric
formula is misleading and it is virtually trivial to verify that
$\Theta=\Theta^\dagger$ as required.

We should add that as long as the less economical, double-series
formula (\ref{reforum}) for the metric is concerned, it might
still find some marginal applications in some more complicated
quantum toboggans \cite{tobo} etc. Moreover, its use may always be
combined with our present, single-series recipe. For example, in
comparison with the recommendations formulated in ref.~\cite{I},
the necessary nondiagonal matrix coefficients are now much more
easily evaluated,
 \be
  M_{\lambda,\lambda'} =
 \br\,\lambda\,|\,\Theta\,|\,\lambda'\kt
 =
 \bbr\,\lambda\,|\,\lambda'\kt=
 \brz\,\lambda\,|W^{-1}|\,\lambda'\,\kt =
 \bbr\,\lambda\,|W^{-1}|\,\lambda'\,\}\,.
  \label{reforumde}
 \ee
Moreover, the single-series expansion (\ref{neebe}) of the
Sturmian metric could prove too compact when the analysis of some
symmetries is concerned.

Of course, the advantages of our present single-series formula
will prevail in the majority of applications where, typically,
people truncate the infinite series in order to obtain a
reasonable approximation of observable quantities \cite{Batal},
etc. Moreover, even without immediate numerical applications, the
existence and compact form of such a formula can definitely be
seen as an important indication of the mathematical as well as
physical consistency and tractability of the whole family of
non-Hermitian, cryptohermitian Sturmian models.

\section{Summary \label{summ} }

The standard areas of applicability of the general concept of
Sturmians has been reviewed by Rotenberg \cite{Rotenberg}. The
scope of this review ranges from the electron-hydrogen scattering
to the interatomic charge-transfer collisions and to the
atomic-physics related solutions of the Faddeev and
Born-Oppenheimer dynamical and phenomenological equations. In the
not too remote area of mathematical physics the transition from
bound states [cf. eq.~(\ref{SErovni})] to Sturmians  [cf.
eq.~(\ref{SErovnice})] found typical applications in perturbation
theory \cite{rnad}.

Besides such a role aiming at an immediate determination of
energy-dependent couplings the phenomenological use of Sturmians
involves mathematically motivated constructions of the so called
quasi-exactly solvable quantum models~\cite{Voros}. The latter
models with $W(r)\sim r^M$ at positive (half)integer $M$
re-attracted attention to many open questions like, e.g., the
completeness of the sets of the Sturmian wave functions
\cite{refe}. Last but not least, one should keep in mind that
certain exact Sturmians emerging at $W(r)=r^N$ with any $N = -1,
0, 1, 2, \ldots$ proved helpful, in the context of classical
physics, in connection with the resonant internal boundary layer
problem \cite{BW}.

Our preceding paper I revisited the problem of  Sturmians and
extended the domain of their applicability to the models
characterized by a loss of manifest Hermiticity of the
Hamiltonian. It has been established there that there exists a
close connection between the concepts of Sturmians and of the so
called crypto-Hermitian Hamiltonians with real spectra. We
emphasized there that the survival of the reality of the energy
spectrum can be expected to occur not only in the non-Sturmian
regime with $W = I$ but also in the presence of the nontrivial
weight operators $W \neq I$.

In our present continuation of paper I we felt inspired by the
lasting interest in the applications of Sturmians mediated by the
problems ranging from the very standard computational physics
\cite{Kelbert} to the rather exotic tobogganic models using  a
fairly abstract concept of complexified coordinates \cite{I,tobo}.
In this setting we perceive paper I as an introduction to the
subject which, naturally, suffered from an incomplete
understanding of many subtleties.

One of these subtleties has been clarified in our present paper.
Our main conclusion is that the parallels between Schr\"{o}dinger
equations of the respective types (\ref{SErovni}) and
(\ref{SErovnice}) are much closer than expected in paper I. In
particular, we feel pleased by the fact that in contrast to the
very sceptical expectations expressed in paper I we succeeded here
in finding the entirely general formula for the metric $\Theta$
and in showing that this operator of key relevance can be
expressed as a single infinite sum over certain elementary
projectors at $W=I$ as well as at $W\neq I$, i.e., in both the
non-Sturmian and Sturmian cases, respectively.

\subsection*{Acknowledgement}

Participation of MZ supported by the M\v{S}MT ``Doppler Institute"
project Nr. LC06002, by GA\v{C}R, grant Nr. 202/07/1307 and, last
but not least, by the hospitality of NiTheP and STIAS in
Stellenbosch.

\end{document}